\documentstyle[aas2pp4]{article}  
\lefthead{Fassnacht \& Cohen}
\righthead{}
\begin{document}

\title{Keck Spectroscopy of Three Gravitational Lens Systems Discovered
in the JVAS and CLASS Surveys}

\author{\sc Christopher D. Fassnacht \& Judith G. Cohen}
\affil{Palomar Observatory, California Institute of Technology, 105-24, 
Pasadena, CA 91125 \\
{\rm cdf@astro.caltech.edu, jlc@astro.caltech.edu}}

\begin{abstract}
We present spectra of three gravitational lens systems taken with the
Low Resolution Imaging Spectrograph on the W.~M.\ Keck Telescopes.
All of the systems were discovered in the JVAS and CLASS radio
surveys, which were designed to find lenses suitable for measuring
$H_0$.  Previous spectra of these systems had low signal-to-noise
ratios, and only one of the source redshifts was secure.  Our
observations give unambiguous lens and source redshifts for all of the
systems, with ($z_{\ell}$,$z_s$) $=$ (0.4060,1.339), (0.5990,1.535)
and (0.4144,1.589) for B0712+472, B1030+074 and B1600+434,
respectively.  The observed image splittings in the systems imply that
the masses of the lensing galaxies within their Einstein rings are
5.4$\times 10^{10}$, 1.2$\times 10^{11}$, and 6.3$\times 10^{10}
h^{-1} M_{\sun}$.  The resulting $V$-band mass-to-light ratios for
B0712+472 and B1030+074, measured inside their Einstein ring radii, are
$\sim 10h~(M/L)_{\sun, V}$, slightly higher than values observed
in nearby ellipticals.  For B1600+434, the mass-to-light ratio is
$48h~(M/L)_{\sun, V}$.  This high value can be explained, at least in
part, by the prominent dust lane running through the galaxy.  Two of
the three lens systems show evidence of variability, so monitoring may
yield a time delay and thus a measurement of $H_0$.

\end{abstract}


\section{Introduction}

It has been known for many years that gravitational lens systems can
in principle be used to determine the values of various cosmological
parameters.  Refsdal (1964) described a method by which variations in
intensity of a multiply-imaged background source could be used to
determine the Hubble Constant.  A well-constrained model of the
lensing potential can be used to predict differences in light travel
time along the multiple paths from source to observer; knowing the
redshifts of the background source and the lensing galaxy for each
system is crucial to this method.  The ratio of the observed to the
predicted time delays is directly proportional to $H_0$.  Time delays
have been measured for two lens systems to date: 0957+561
(\cite{tk09571}; \cite{tk09572}; \cite{ao0957}) and 1115+080
(\cite{s1115}; \cite{bk1115}).  Combining the measured time delays
with current models for these systems has given values of $H_0$ in the
range $\approx\ $60--70~km/sec/Mpc for 0957+561 (\cite{tk09572};
\cite{ao0957}; \cite{ef0957}) and $\approx\ $40--55~km/sec/Mpc for
1115+080 (\cite{s1115}; \cite{kk1115}; \cite{tk1115}; \cite{fc1115}).
Improvements in the time delay measurements and additional constraints
on the lens models are expected.  The value of Refsdal's method will
be proved if several lenses can be used to measure $H_0$ and give
consistent values.

The Jodrell-VLA Astrometric Survey (JVAS; \cite{jvas1}; \cite{jvas2})
and the Cosmic Lens All-Sky Survey (CLASS; \cite{j1600}; \cite{m1608};
\cite{iwbclass}; \cite{stmclass}) contain observations of $\sim$10,000
flat-spectrum radio sources.  One of the primary goals of these
surveys is to find gravitational lens systems which can be used to
measure $H_0$.  To date, 11 new gravitational lenses have been
discovered in the JVAS and CLASS surveys, and 16 lens candidates are
being investigated further.  The lens and source redshifts have been
determined for almost all of the newly discovered lens systems.  This
paper presents spectra taken at the W.\ M.\ Keck Observatory of three
of the systems with missing redshifts, B0712+472, B1030+074 and
B1600+434.  There is evidence of variability in B0712+472
(\cite{j0712}) and B1600+434 (\cite{1600jandh}), so at least two of
the systems present the possibility of being used to measure $H_0$.

We use $H_0 = 100\ h\ {\rm km\ s}^{-1}\ {\rm
Mpc}^{-1}$ and assume $q_0 = 0.5$\ throughout this paper.

\section{Targets}

Below we present information on previous observations of the lens
systems that are discussed in this paper.

\subsection{B0712+472}

This lens system consists of four images of the background source in a
typical lensing geometry (e.g. \cite{bn}).  The maximum image
separation is 1\farcs27.  Radio maps made with the VLA and MERLIN do
not resolve the four images (\cite{j0712}).  Images taken with WFPC2
on HST show all four images and, in addition, the lensing galaxy
(\cite{hstlens}).  Photometry derived from the WFPC2 images give the
lens magnitude as $V \sim 20.2$ and $I \sim 20.0$\footnote{In this
paper all WFPC2 F555W and F814W magnitudes have been converted to
Johnson $V$ and Cousins $I$ magnitudes, respectively, using
transformations in Holtzman et al.\ (1995).}, measured in an
elliptical aperture with major and minor axes of 2 and 1\arcsec,
respectively (\cite{j0712}).  The total source magnitudes, assuming
the images were point sources, were $V \sim 23$ and $I \sim 22.5$.
Low signal-to-noise ratio optical spectra of the system, taken with
the William Herschel Telescope (WHT), show weak broad emission lines
(\ion{C}{4}, \ion{C}{3}], \ion{Mg}{2}) giving a source redshift of
$z_s \sim 1.33$, and a ``hint of Ca H and K'' absorption, which
tentatively gives $z_{\ell} \sim 0.40$ (\cite{j0712}).  The redshifts
derived from the WHT spectra, especially that of the lens, are not
secure.  However, both are confirmed to be correct by the observations
described in this paper.

\subsection{B1030+074}

The B1030+074 system consists of two flat-spectrum radio components
separated by 1\farcs56 with a flux ratio of 15 to 1 (\cite{iwbclass};
\cite{x1030}).  These components are unresolved at the milliarcsecond
resolution of VLBI.  The system was imaged with WFPC2 and both images
of the background source are seen, as is the lensing galaxy.
Component A, the brighter image, has magnitudes of $V \sim 20$\ and $I
\sim 19$ and component B is $\sim$2.5 -- 3 magnitudes fainter in each
band; the lensing galaxy is estimated to have magnitudes $I \sim
20.5$\ and $V \sim 22$ (\cite{x1030}).  Spectra taken with the COSMIC
spectrograph (\cite{cosmic}) on the Hale Telescope show no clear
spectral features (Vermeulen \& Womble, private communication).

\subsection{B1600+434}

This system is also a double, with separation 1\farcs4 and a flux
ratio of 1.30 to 1 at radio wavelengths (\cite{j1600}).  Jaunsen \&
Hjorth (1997) have imaged the system with the Nordic Optical
Telescope and find that the lensing galaxy is an edge-on spiral with
$B = 23.6, V = 22.0, R=21.1$, and $I = 20.3$.  The background source
images have total magnitudes of $B = 21.9, V = 21.7, R=21.1$, and $I =
20.4$.  Imaging with WFPC2 has shown that the lensing galaxy has a
prominent dust lane along its major axis (\cite{hstlens}).  Previous
spectroscopy with the WHT detected broad \ion{C}{4}, \ion{C}{3}] and
\ion{Mg}{2} emission lines from the background source.  These lines
give a source redshift of $z_s = 1.61$, but no redshift for the lens
was determined (\cite{j1600}). Jaunsen \& Hjorth (1997) used the lensing
galaxy colors to get a photometric redshift of $z \sim 0.4$, which is
confirmed by the observations presented in this paper.

\section{Observations and Data Reduction}

All three of the systems were observed with the Low Resolution Imaging
Spectrograph (LRIS; \cite{lris}) in longslit mode on the Keck
Telescopes.  The 300 g/mm grating was used, giving a scale of
2.44~\AA/pix.  The longslit was aligned with the major axis of the
lensing galaxy (B0712+472 and B1600+434) or the axis defined by the
two images of the background source (B1030+074). All objects were
observed for 3000~sec total integration time; other details of the
observations are given in Table~\ref{tab_obs}.  The data were reduced
using standard IRAF\footnote{IRAF (Image Reduction and Analysis
Facility) is distributed by the National Optical Astronomy
Observatories, which are operated by the Association of Universities
for Research in Astronomy (AURA) under cooperative agreement with the
NSF.} routines.  The bias levels were estimated using the overscan
region on the chip.  For the observations of B1600+434, the flat-field
frame was constructed from dome flats; in all other cases, exposures
with the internal flat-field lamp were used.  The spectra were
extracted using the IRAF implementation of the ``optimal extraction''
technique described in Horne (1986) and Marsh (1989).  For the 1997
Feb 07 observation of B0712+472 and the 1997 Feb 14 observation of
B1030+074, the emission from the background source and the lensing
galaxy were spatially separated on the slit, so two spectra were
extracted for each system.  Wavelength calibration was performed using
sky lines (1997 Feb 07) or arc lamps taken after each science exposure
(all other observations).  Observations of the Oke spectrophotometric
standard stars G191B2B and HZ44 (\cite{okestds}) were used to remove
the response function of the chip.  For the data taken on 1996 June 18
and 1997 Feb 14, exposures of BL~Lac objects were used to remove
atmospheric absorption features.  The individual spectra for each
object were weighted by their signal-to-noise ratios and combined.

\begin{deluxetable}{lllccrc}
\tablewidth{0pt}
\scriptsize
\tablecaption{Observations\label{tab_obs}}
\tablehead{
   \colhead{Source}
 & \colhead{Date}
 & \colhead{Telescope}
 & \colhead{$t_{exp}$ (sec)}
 & \colhead{Slit Width (\arcsec)}
 & \colhead{P.A.}
 & \colhead{Coverage (\AA)}
}

\startdata

B0712+472 & 1997 Feb 06 & Keck II & 3000 & 0.7 & 75     & 3593--8588 \\
          & 1997 Feb 14 & Keck II & 3000 & 0.7 & 79     & 4208--9206 \\
B1030+074 & 1997 Feb 14 & Keck II & 3000 & 1.0 & 322    & 4211--9208 \\
B1600+434 & 1996 Jun 18 & Keck I  & 3000 & 1.0 & $-120$ & 4606--9358 \\

\enddata
\end{deluxetable}

\begin{figure}
\plotone{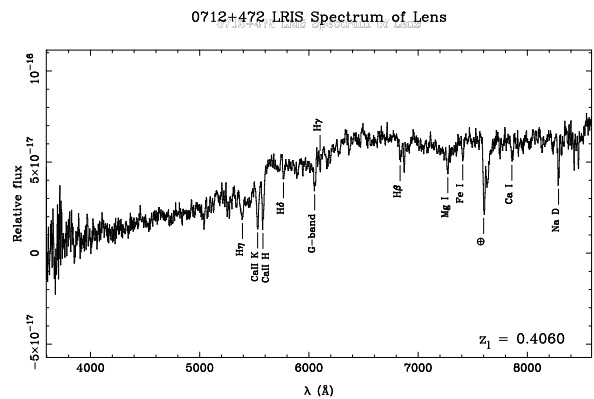}
\caption{\label{0712lspec} Spectrum of lensing galaxy in B0712+472 system taken
on 1997 Feb 06.}
\end{figure}

\begin{figure}
\plotone{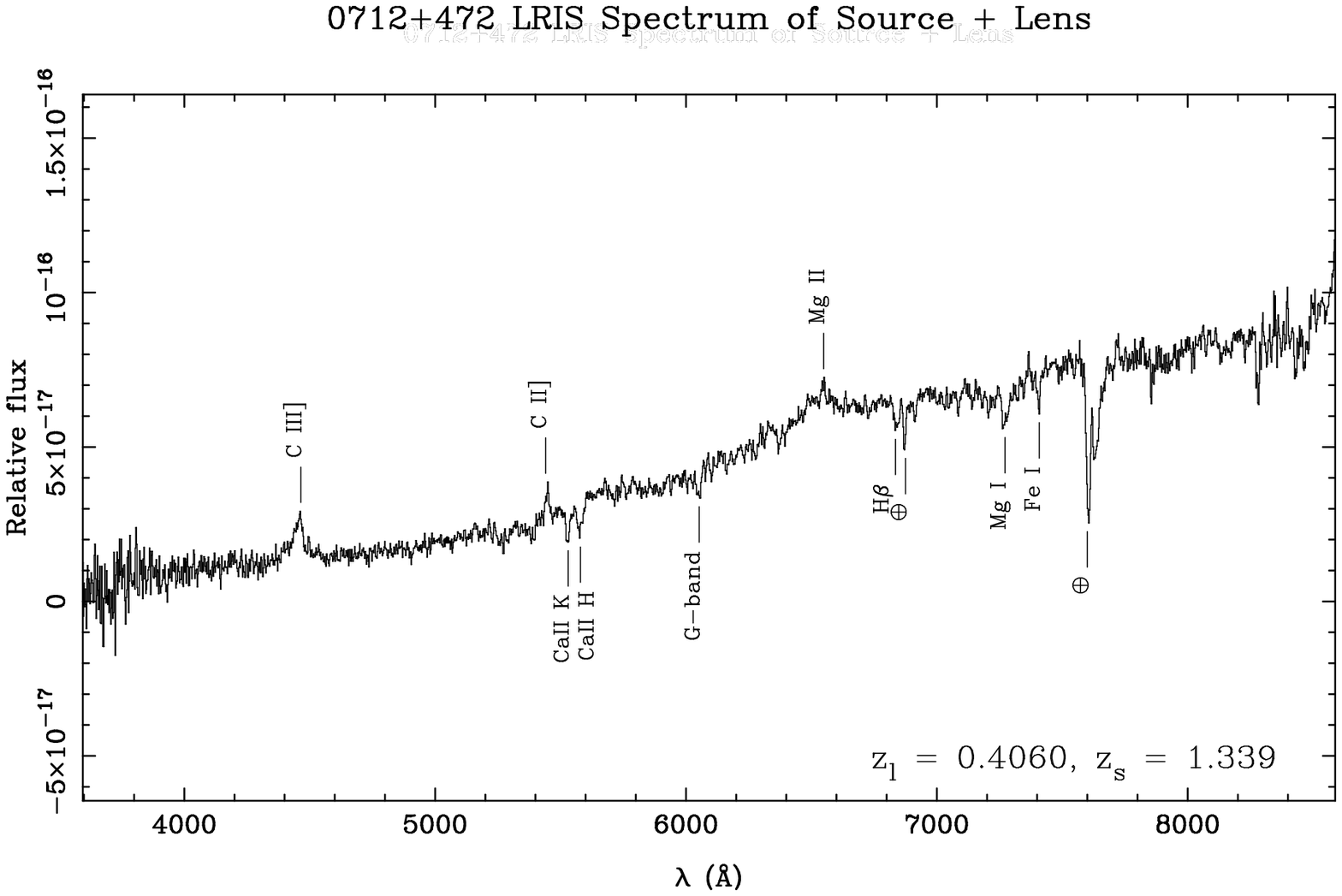}
\caption{\label{0712sspec} Spectrum of background source in B0712+472
system taken on 1997 Feb 06.  The spectrum also includes a significant
amount of light from the extended lensing galaxy.  The \ion{C}{3}],
\ion{C}{2}] and \ion{Mg}{2} are associated with the background source.
All other lines are associated with the lensing galaxy.}
\end{figure}

\begin{figure}
\plotone{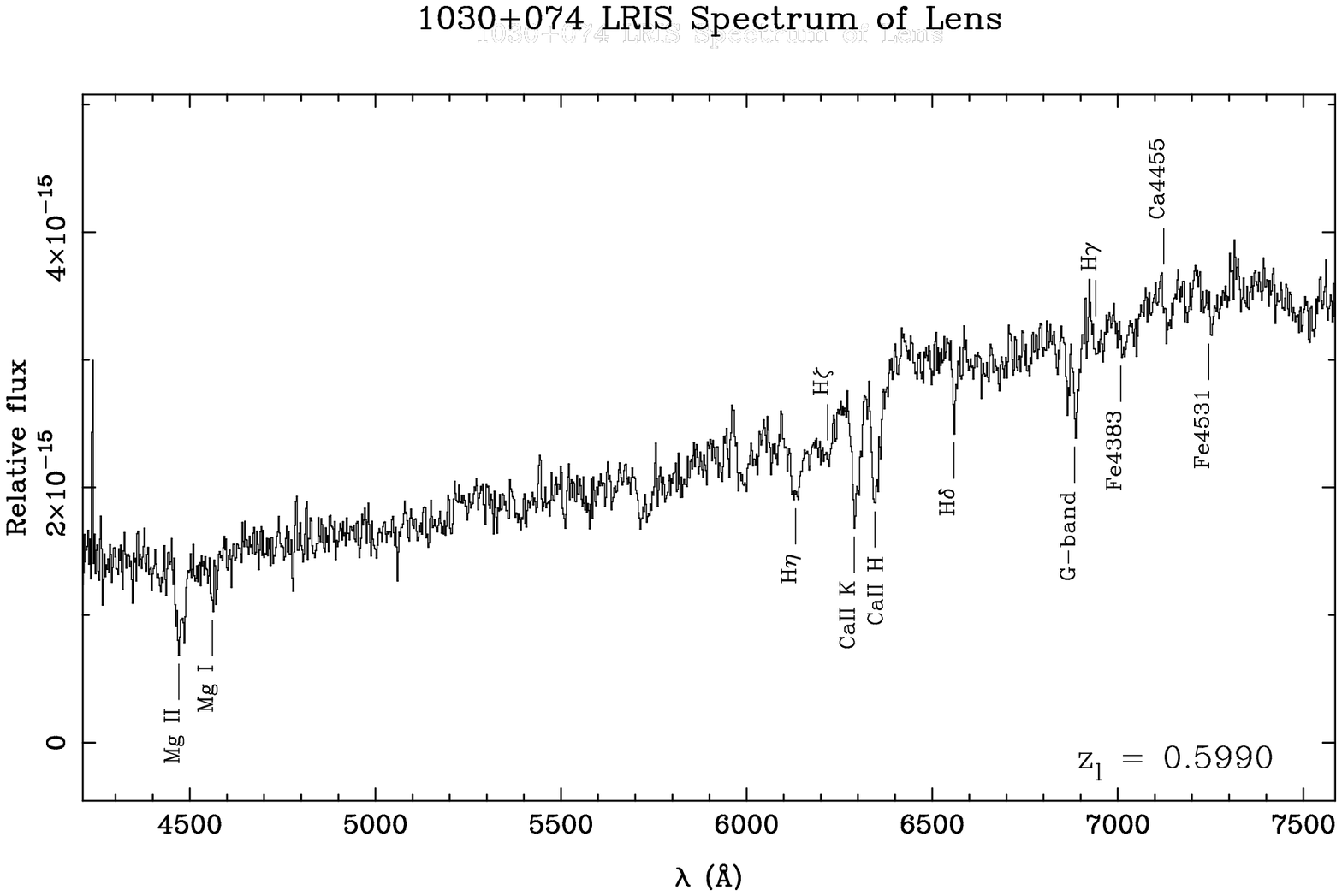}
\caption{\label{1030lspec} Spectrum of lensing galaxy in B1030+074 system taken
on 1997 Feb 14.}
\end{figure}

\begin{figure}
\plotone{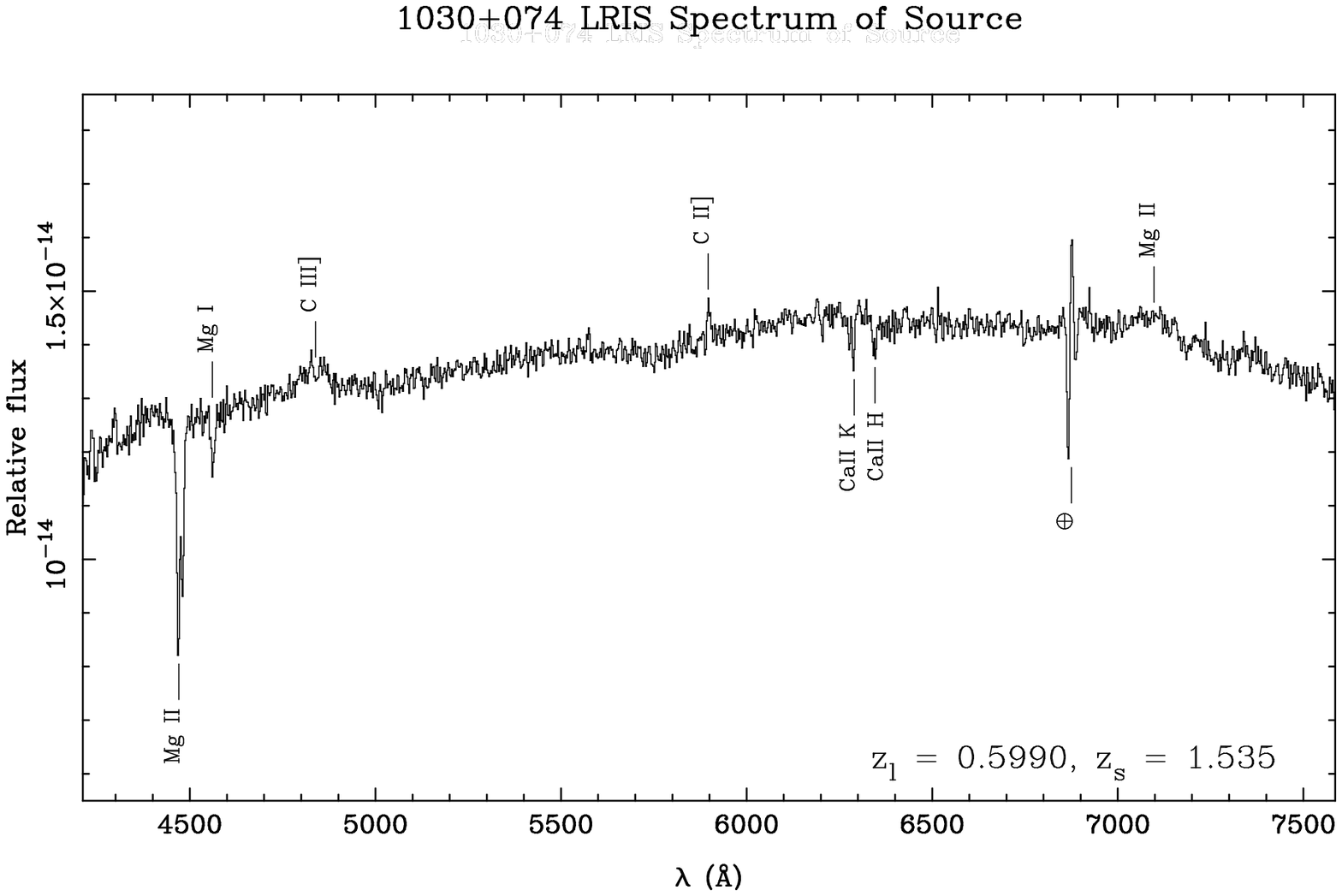}
\caption{\label{1030sspec} Spectrum of background source in B1030+074 system
taken on 1997 Feb 14.  The \ion{C}{3}], \ion{C}{2}] and \ion{Mg}{2} emission
lines are
associated with the background source.  All other lines are associated
with the lensing galaxy.}
\end{figure}

\begin{figure}
\plotone{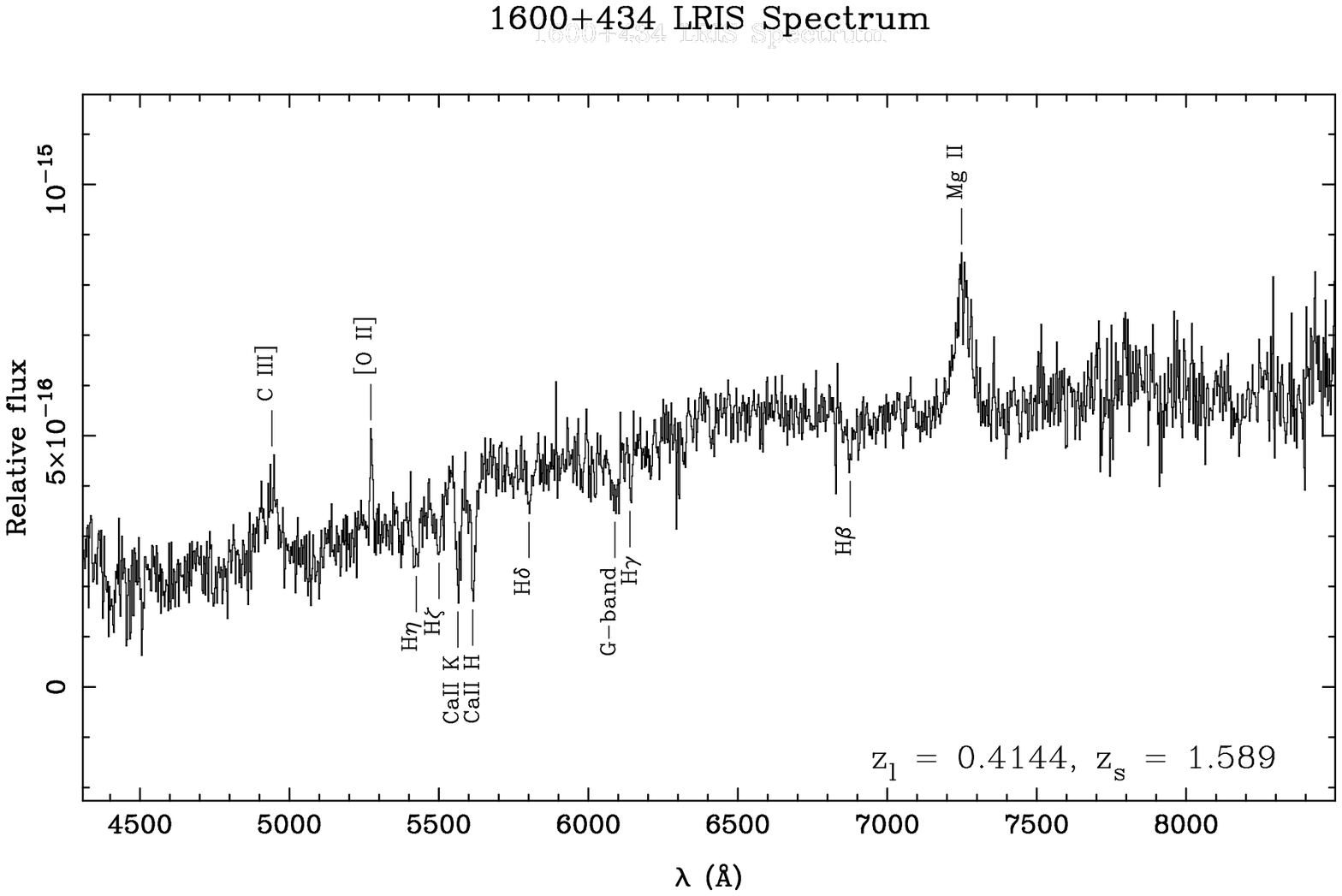}
\caption{\label{1600spec} Spectrum of B1600+434 containing light from both
the lens and the source.  The \ion{C}{3}] and
\ion{Mg}{2} emission lines are associated with the background source.  All
other lines are associated with the lensing galaxy.}
\end{figure}

\section{Results}

The final spectra are shown in Figures \ref{0712lspec} --
\ref{1600spec}.  For each lens system, the spectral features
(Table~\ref{tab_spec}) can be explained as the sum of a lens at
$z_{\ell}$\ and a background source at $z_s$, where a single choice of
$z_{\ell}$\ and $z_s$\ (Table~\ref{tab_pars}) suffices.  Uncertainties
in the redshifts were estimated by taking the RMS scatter in the
redshifts calculated from the individual spectral lines.  For B1600+434
the source redshift uncertainty was estimated as half the difference
of the redshifts calculated from the two broad emission lines.
Further discussion of the individual systems follows below.

\subsection{B0712+472}

The B0712+472 spectra show a lensing galaxy at $z_{\ell}=0.4060$ and a
background source at $z_s=1.339$.  These results confirm the Jackson
et al.\ (1997b) tentative redshifts.  The lensing galaxy has a 
typical early-type galaxy spectrum (e.g.\ \cite{galspecs}).  The
moderately strong 4000\AA\ break, small equivalent-width Balmer
absorption lines and lack of \ion{O}{2} emission indicate that little
star formation is occuring in this galaxy.  The background source
shows broad \ion{C}{3}] and \ion{Mg}{2} emission lines typical of a
quasar spectrum.

\subsection{B1030+074}

The B1030+074 system contains a lensing object at $z_{\ell} = 0.5990$\
and a background source at $z_s = 1.535$.  In this system as well, the
lensing object spectrum is typical of an early-type galaxy.  The
background source shows broad \ion{C}{3}] and \ion{Mg}{2} emission
lines.  The broad emission lines have low equivalent widths, raising
the possibility that the background source is a BL Lac-type object.
It should be noted, however, that the source spectrum is almost
certainly contaminated by some light from the lensing object, which
will raise the continuum level and reduce the apparent equivalent
widths of the lines.

\subsection{B1600+434}

The B1600+434 spectrum has the lowest signal-to-noise ratio of the
observations presented in this paper.  However, both lens and source
redshifts ($z_{\ell}=0.4144, z_s = 1.589$) are clearly determined.
The background source is a quasar with broad \ion{C}{3}] and
\ion{Mg}{2} emission lines, as was seen in Jackson et al.\ (1995).
The lens spectrum is indicative of a later galaxy type than that
observed in the B0712+472 and B1030+074 systems.  This spectrum has
[\ion{O}{2}] emission and a less prominent 4000\AA\ break.  This is
consistent with images of the lensing galaxy which show that it is
an edge-on spiral (\cite{1600jandh}; \cite{hstlens}).  For later
calculations, we will assume that the lensing object is an Sa or Sb
galaxy.  However, determining the galaxy type from the spectrum may be
complicated by extinction from the dust lane, which runs the length of
the observable disk (\cite{hstlens}).

\section{Discussion}

In order to use a lens system to measure $H_0$, it is necessary
to first know the source and lens redshifts.  These redshifts are
used to derive the angular diameter distances to the source and
the lens, which have an inverse dependence on $H_0$.
The observed image positions and other information can then be used
to construct a model of the lensing potential.  The lens model is
combined with the angular diameter distances to give predicted time
delays:
$$\Delta t_i = (1 + z_{\ell})\frac{D_{\ell} D_S}{c D_{\ell s}}
   \left[\frac{1}{2} |\theta_i - \beta|^2 - \phi(\theta) \right],$$
where $D_{\ell}$, $D_S$ and $D_{\ell s}$ are the angular diameter
distances to the lens, to the source, and between the lens and
source, respectively; $\theta_i$ is the image position; $\beta$ is
the source position and $\phi$ is the scaled lensing potential
(e.g.\ \cite{hogg}).  The predicted time delay is proportional
to $h^{-1}$ from the ratio of angular diameter distances.  Thus, if the
background source is variable and time delays can be measured, the
ratio between the observed and predicted delays will give $h$.

We can also use these lenses to study the properties of galaxies at
moderate redshifts.  For example, the extent of the image-splitting by
the lens gives a dircect estimate of the mass inside the Einstein ring
of the lens.  For the lens systems discussed in this paper, the
Einstein ring radii correspond to physical sizes of 2 --
3$h^{-1}$~kpc.  The mass is estimated as (e.g.\ \cite{bn}):
$$M_E \sim 1 \times 10^{12} \left(\frac{D}{1~{\rm Gpc}}\right)
\left(\frac{\theta_E}{3~\arcsec}\right)^2 M_{\sun}, \quad D \equiv
\frac{D_{\ell}D_s}{D_{\ell s}},$$ where $\theta_E$ is the angular
radius of the Einstein ring.  For these lenses, we find values of $M_E
\sim 5 \times 10^{10} - 1 \times 10^{11} h^{-1} M_{\sun}$.  

With the above masses and the photometry for these systems, we can
compute approximate mass-to-light ratios for the lensing galaxies.
For B0712+472 and B1030+074, the lens magnitudes discussed in Section
2 were taken within apertures roughly the size of the Einstein rings
in the systems, so we use those values in the following calculations.
For B1600+434 we have used the WFPC2 images to estimate the lens
magnitude in the appropriate aperture, finding $I \sim 21.8$.  We
converted the observed $I$-band magnitudes to rest-frame $V$-band
magnitudes using $k$-corrections for the lens redshifts and rest-frame
$(V - I)$ colors for typical E/S0 (for B0712+472 and B1030+074) and Sa
(B1600+434) galaxies.  The resulting mass-to-light ratios for
B0712+472 and B1030+074 are $\sim 10 h (M/L)_{\sun , V}$.  For
B1600+434 we find $(M/L)_E = 48 h (M/L)_{\sun , V}$, confirming the
value found by Jaunsen \& Hjorth (1997).

The values of mass-to-light for B0712+472 and B1030+074 are slightly
higher than mass-to-light ratios of nearby ellipticals within the same
physical radii (e.g. \cite{mtole_vdm}; {\cite{mtole_g}).  We note that
we are biased toward finding high mass systems when looking for
gravitational lenses, since these systems have a larger cross-section
for lensing.  Also, the slightly different apertures used to compute
the masses and luminosities may be biasing the results toward higher
mass-to-light ratios.  In addition, the presence of dust in the
lensing galaxies would cause some of the light to be lost, thereby
increasing the observed mass-to-light ratio.  The dust lane in
B1600+434 is clearly seen and is responsible, at least in part, for
the very high mass-to-light ratio calculated for this system.  There
is also evidence for absorption by dust in B0712+472 (\cite{hstlens}).
Evidence for dust has been seen in other lens systems as well
(\cite{dustylens}; \cite{malhotra}), indicating that dusty lenses may
be quite common.  Infrared imaging of these systems would provide more
accurate measurements of the mass-to-light ratios.

\section{Prospects for Measuring $H_0$}

In order to use these systems to measure $H_0$, well-constrained
models of the lensing potentials must be constructed.  The following
observations could provide data that would aid in the modelling: (1)
detecting milliarcsecond-scale structure in the radio images, (2)
determining the velocity dispersion of the lensing galaxy or, in the
case of B1600+434, measuring the galaxy rotation curve, and (3) imaging
the systems at high-resolution with NICMOS to provide accurate
positions of the lensed images with respect to the lensing galaxy,
unbiased by dust extinction.

The final piece of the $H_0$ puzzle is measuring time delays between
the lensed images.  Jackson et al.\ (1997b) find slight variability in
the component fluxes in B0712+472, with a $\sim$30\% overall decrease
in flux between 1995 and 1996.  The background source in B1600+434
appears to be variable as well.  Our VLA observations of B1600+434
show that between 1995 and 1996, the flux densities of the two
components decreased by $\sim$50\%, and the component flux ratio
changed from 1.30 to 1.11.  In addition, Jaunsen and Hjorth (1997) see
optical variations in B1600+434. Thus, at least two of the sources
show variability and, if observed regularly, present the possibility
of being used to measure $H_0$.

\begin{deluxetable}{lclccc}
\tablewidth{0pt}
\scriptsize
\tablecaption{Detected Spectral Lines\label{tab_spec}}
\tablehead{
 &
 & 
 & \multicolumn{3}{c}{Observed Wavelength} \\
   \colhead{Ion}
 & \colhead{$\lambda_0$ (\AA)}
 & \colhead{Object}
 & \colhead{B0712+472}
 & \colhead{B1030+074}
 & \colhead{B1600+434}
}

\startdata
\ion{C}{3}]   & 1909 & Source & 4461    & 4834    & 4930 \\
\ion{C}{2}]   & 2326 & Source & 5448    & 5898    & \nodata \\
\ion{Mg}{2}   & 2796 & Source & 6540    & 7095    & 7252 \\
\\
\ion{Mg}{2}   & 2796 & Lens   & \nodata & 4469    & \nodata \\
\ion{Mg}{2}   & 2802 & Lens   & \nodata & 4480    & \nodata \\
\ion{Mg}{1}   & 2852 & Lens   & \nodata & 4563    & \nodata \\
\ion{O}{2}    & 3727 & Lens   & \nodata & 5963    & 5274 \\
H$\eta$       & 3835 & Lens   & 5391    & 6134    & 5418 \\
H$\zeta$      & 3889 & Lens   & \nodata & 6212    & 5498 \\
\ion{Ca}{2} K & 3934 & Lens   & 5529    & 6288    & 5566 \\
\ion{Ca}{2} H\tablenotemark{a} & 3968 & Lens   & 5579    & 6345    & 5615 \\
H$\delta$     & 4102 & Lens   & 5768    & 6560    & 5802 \\
G-band        & 4300 & Lens   & 6053    & 6887    & 6094 \\
H$\gamma$     & 4340 & Lens   & 6104    & 6943    & 6143 \\
H$\beta$      & 4861 & Lens   & 6839    & \nodata & 6876 \\
\ion{Mg}{1}   & 5172 & Lens   & 7266    & \nodata & \nodata \\
\ion{Mg}{1}   & 5183 & Lens   & 7287    & \nodata & \nodata \\
\ion{Fe}{1}   & 5269 & Lens   & 7409    & \nodata & \nodata \\
\ion{Ca}{1}   & 5589 & Lens   & 7863    & \nodata & \nodata \\
\ion{Na}{1} D & 5890 & Lens   & 8283    & \nodata & \nodata \\
\ion{Na}{1} D & 5896 & Lens   & 8290    & \nodata & \nodata \\

\enddata
\tablenotetext{a}{May be blended with H$\epsilon$\ $\lambda$3970.}
\end{deluxetable}

\begin{deluxetable}{lclccccc}
\tablewidth{0pt}
\scriptsize
\tablecaption{Lens System Parameters\label{tab_pars}}
\tablehead{
   \colhead{System}
 & \colhead{$z_{\ell}$}
 & \colhead{$z_s$}
 & \colhead{$D_{\ell}$}
 & \colhead{$D_s$}
 & \colhead{$D_{\ell s}$}
 & \colhead{$M_E$\tablenotemark{a}}
 & \colhead{$(M/L)_V$\tablenotemark{b}}
\\
 &
 &
 & \colhead{$(h^{-1}~{\rm Mpc})$}
 & \colhead{$(h^{-1}~{\rm Mpc})$}
 & \colhead{$(h^{-1}~{\rm Mpc})$}
 & \colhead{$(h^{-1}~M_{\sun})$}
 & \colhead{$(h~(M/L)_{\sun})$}
}

\startdata
B0712+472 & 0.4060 $\pm$ 0.0002  & 1.339 $\pm$ 0.002 & 668 $\pm$ 0.16  
          & 887 $\pm$ 0.042 & 486 $\pm$ 0.34 & 5.4 $\pm 0.22 \times 10^{10}$ 
          & 8.6 $\pm$ 0.93 \\
B1030+074 & 0.5990 $\pm$ 0.0003  & 1.535 $\pm$ 0.003 & 784 $\pm$ 0.13  
          & 879 $\pm$ 0.16  & 385 $\pm$ 0.46 & 1.2 $\pm 0.050 \times 10^{11}$ 
          & 11 $\pm$ 1.2 \\
B1600+434 & 0.4144 $\pm$ 0.0003  & 1.589 $\pm$ 0.006 & 675 $\pm$ 0.24  
          & 876 $\pm$ 0.36  & 507 $\pm$ 0.53 & 6.3 $\pm 0.25 \times 10^{10}$ 
          & 48 $\pm$ 5.2 \\

\enddata
\tablenotetext{a}{Uncertainties calculated assuming a 1\% uncertainty in image 
separation.}
\tablenotetext{b}{Uncertainties calculated assuming a 0.1 magnitude uncertainty 
in photometry.}
\end{deluxetable}

\acknowledgments 

We thank Nicole Vogt, Tony Readhead and the anonymous referee for
insightful comments on the manuscript, and are grateful to Lori Lubin,
Mike Pahre and David Hogg for many helpful discussions.  We are
indebted to Terry Stickel, Wayne Wack, Chuck Sorenson and the Keck
staff for assistance with the observations.  The W.~M. Keck
Observatory was made possible by a generous grant from the W.~M.\ Keck
Foundation.  This work is supported by the NSF under grant \#AST
9420018.

\end{document}